\documentclass[sigconf,9pt]{acmart}

\usepackage{booktabs} 
\usepackage{url}
\usepackage{svg}
\usepackage{etoolbox}
\usepackage{cuted}
\usepackage{lipsum}
\usepackage{booktabs} 
\usepackage{url}
\usepackage{svg}
\usepackage{etoolbox}
\usepackage{cuted}
\usepackage{lipsum}
\usepackage{amssymb}
\usepackage{array}
\usepackage{amsmath}
\usepackage{tikz}
\usepackage{color}
\usepackage{dirtytalk}
\usepackage{algpseudocode}
\usepackage{algorithm}
\usepackage{graphicx}
\usepackage{caption}
\usepackage{mdframed}
\usepackage{tcolorbox}

\copyrightyear{2019} 
\acmYear{2019} 
\setcopyright{acmcopyright}
\acmConference[ICBCT 2019]{2019 International Conference on Blockchain Technology (ICBCT 2019)}{March 15--18, 2019}{Honolulu, HI, USA}
\acmPrice{15.00}
\acmDOI{10.1145/3320154.3320164}
\acmISBN{978-1-4503-6268-9/19/03}

\begin{document}


\title{Multi-Authority Attribute-Based \\ Access Control with Smart Contract}

\author{Hao Guo~~~~~~~~~~\quad Ehsan Meamari~~~~~~~~~ \quad Chien-Chung Shen}
\affiliation{%
  \institution{Department of Computer and Information Sciences, University of Delaware, U.S.A.}
}
\email{{haoguo,ehsan,cshen}@udel.edu}


\begin{abstract}
Attribute-based access control makes access control decisions based on the assigned attributes of subjects and the access policies to protect objects by mediating operations from the subjects. Authority, which validates attributes of subjects, is one key component to facilitate attribute-based access control. In an increasingly decentralized society, multiple attributes possessed by subjects may need to be validated by multiple different authorities.
This paper proposes a multi-authority attribute-based access control scheme by using Ethereum's smart contracts. 
In the proposed scheme, Ethereum smart contracts are created to define the interactions between data owner, data user, and multiple attribute authorities. A data user presents its attributes to different attribute authorities, and after successful validation of attributes, obtains attribute tokens from respective attribute authorities. After collecting enough attribute tokens, a smart contract will be executed to issue secret key to the data user to access the requested object. The smart contracts for multi-authority attribute-based access control have been prototyped
in Solidity, and their performance has been evaluated on the Rinkeby Ethereum Testnet. 
\end{abstract}

%
%



\maketitle

\section{Introduction}

In computer and information security, the mechanism of access control has been widely used to protect objects from unauthorized operations by subjects.  Such operations may include creating, deleting, discovering, editing, executing, event recording~\cite{guo2018blockchain}, and reading of objects. Formally, access control mechanisms is defined as {\em the logical component that serves to receive the access request from the subject, to decide, and to enforce the access decision} \cite{nist}.
With access control, owners of  objects have the authority to establish access policies that govern which subjects may perform what operations on the objects. How these access policies are specified depends on the access control models within which the subjects, objects, operations, and rules interact to make and enforce access control decisions.
Although each model has its own advantages and limitations, over time, access control models have evolved from identity-based to role-based, and to attribute-based.

In general, attribute-based access control makes access control decisions based on the assigned attributes of subjects and the access policies (complex Boolean rules evaluating attributes) to protect objects by mediating operations from the subjects.
Broadly speaking, attribute-based access control has been defined as {\em an access control method where subject requests to perform operations on objects are granted or denied based on assigned attributes of the subject, assigned attributes of the object, environment conditions, and a set of policies that are specified in terms of those attributes and conditions} \cite{nist}.

To facilitate attribute-based access control, one critical component is an {\em authority} which validates attributes of the subjects to make access control decisions. In theory, one attribute authority (AA) may oversee and validate all the attributes of subjects. However, in an increasingly decentralized society, there may exist multiple authorities which take part in validating different attributes possessed by subjects.  For instance, a Delaware resident is a student of the University of Delaware and also interns at the DuPont company. These three attributes must be validated by three different authorities, the University of Delaware, Delaware's Division of Motor Vehicles, and the DuPont company. As another example of a commercial application, two banks, such as JP Morgan Chase and Bank of America, hence two authorities, may both need to validate respective attributes of people who take part in a joint project.

In this paper, we propose a scheme to facilitate multi-authority attribute-based access control by using Ethereum smart contracts. In this scheme, data owner generates a secret key and encrypt the shared data with the AES algorithm, and keeps the secret key with himself. Within Ethereum's smart contracts, a data user presents attributes to respective authorities to obtain attribute tokens after successful validation. Upon collecting enough  attribute tokens from multiple attribute authorities, the data user will receive the AES secret key capable of accessing the request data. Our contributions are mainly two parts: First, we propose the multi-authority attribute-based access control mechanism by designing smart contracts and utilizing Ethereum blockchain platform. Second, we implement the attribute token rules to represent the attribute sets and access policy in traditional attribute-based access control mechanism. With the design of multiple attribute authorities, we achieve the multi-authority functionality for our scheme. 


The remainder of this paper is organized as follows. In Section \ref{sec:pre}, backgrounds of  Ethereum smart contract and attribute-based access policy are reviewed. Then, the architecture of smart contract-based multi-authority access control scheme, and the detailed designs of smart contracts are presented in Section \ref{sec:arch}. In Section \ref{sec:evaluation}, we evaluate the costs of executing the smart contracts and analysis the security and privacy issues.  Related work is described in Section \ref{sec:related work}, and Section \ref{conclusion} concludes the paper.
\section{Preliminary}
\label{sec:pre}

In this section, we review the basic concepts of Ethereum with smart contract and attribute-based access policy.

\subsection{Ethereum with Smart Contract}

Ethereum is a distributed computing platform  featuring the smart contract functionality~\cite{ethreport},\cite{ethofficial}. It is proposed in late 2013 by Vitalik Buterin, a cryptocurrency researcher and programmer~\cite{ethwikisite}.  The Ethereum development was funded by an online crowdsale event which took place between July and August 2014, and the official system went live on 30 July 2015~\cite{ethwikisite}.  In contrast to Bitcoin-like systems where transactions are programmed in a simple non-Turing complete scripting language, and can only specify basic logic relations, which limits their utility to other application domains, Ethereum smart contract provides an event-driven, Turing complete scripting functionality to specify and process complex transactions which can be verified to demonstrate the feasibility of the contract operation. From the perspective of the smart contract, it works like the a event-driven script and will automatically execute the script if the pre-defined logical condition has been satisfied. Before the smart contract is executed, all related logic functions and processes were already established.  

Within Ethereum, there are two types of accounts: Externally Owned Accounts (EOA) and Contract Accounts, both of which are uniquely identified by a 20-byte hexadecimal string as their address.
An EOA is controlled by its owner's private key, has an available ether balance, and can send transactions (for instance, send a message to another account for transferring ether or trigger the execution of a smart contract). It has no associated code with EOA account. While a contract account also has an ether balance, but contains the associated code which can be triggered by a transaction or from other smart contract.  

\begin{figure}[h]
\centering
\includegraphics[width=0.13\textwidth]{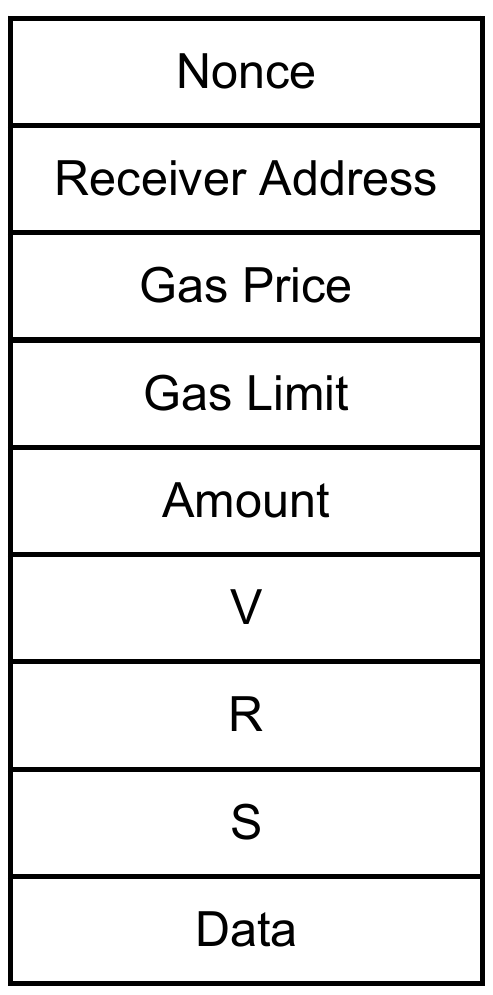}
\caption{Data structure of an Ethereum transaction.}
\label{fig:pic2}
\end{figure} 

Ether is the official cryptocurrency used in Ethereum platform. All the ether balances and values are represented in units of wei: 1 ether is 1e18 wei. The Ethereum platform has a smart contract running environment, which is known as the Ethereum Virtual Machine (EVM)~\cite{cruz2018rbac}. Each mining node in the Ethereum network (Node receives, broadcast, verifies, and executes the transaction in Ethereum network) runs EVM as part of the validation procedure and later perform the same results and store the data. Every operation in EVM has a specific cost, which is counted by the amount of gas. The sender of the transaction needs to pay for ether for the operations and the total transaction cost is estimated as Ether = Gas Used * Gas Price.
\begin{figure}[h]
\centering
\includegraphics[width=0.434\textwidth]{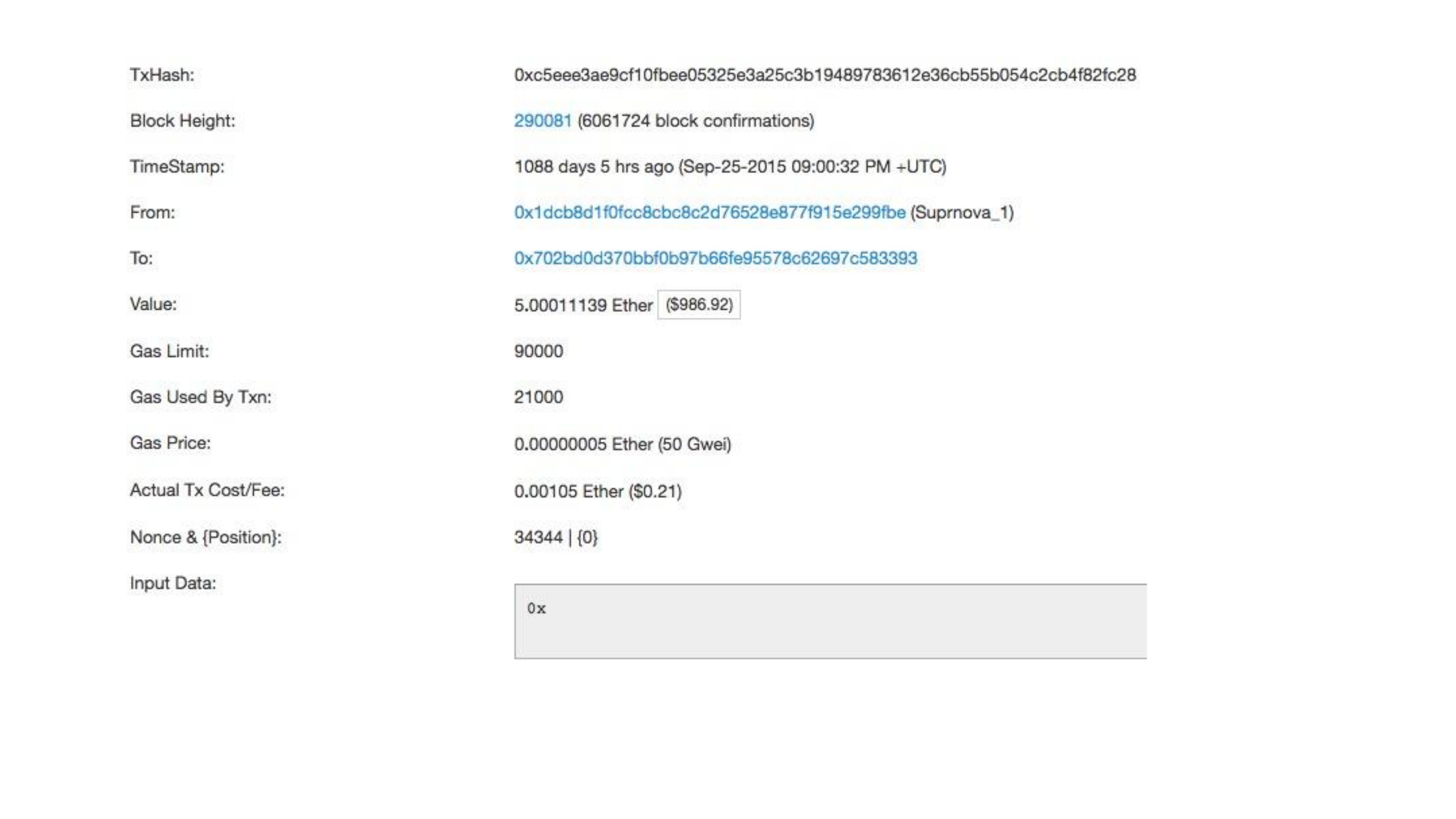}
\caption{Ethereum transaction example from etherscan.io.}
\label{fig:pic2}
\end{figure} 

The Ethereum transaction is a data package which enables user to transfer ether from one account to another account. In addition, it can also trigger the execution of the code in the smart contract through one transaction. Figure 2 and 3 shows the Ethereum transaction data structure and one screenshot from the Etherscan website (etherscan.io). One unique Ethereum transaction consists of the following data field: Nonce, represents for the transaction sequence number from the sender. Receiver address, which contains the receiver account information. Gas price, the price user offer to pay. Gas limit, maximum amount of gas allowed for one transaction. Amount, the total ether balance transfered to the destination address. V, R, and S, together makes up the Elliptic Curve Digital Signature Algorithm (ECDSA) for sender's signature. Data, optional additional data fields which can be put into any data. 

\subsection{Attribute-Based Access Policy}
Attribute-based access control (ABAC) defines an access control policy, in which the access rights are granted to users through the use of access policies which combine attributes together~\cite{wiki:abac}. The access policies could use any type of attributes. For instance, subject's attribute, object's attribute, and environment attributes. Access policy can support the Boolean rule, in which rules contain "IF, THEN" statements about who is making the request, the resource, and the action~\cite{wiki:abac}. For example: IF the requester is a University of Delaware graduate student, THEN allow the access to the data. The critical feature of ABAC is the concept of access policies which can express a complex Boolean rule set that can evaluate many different attributes~\cite{nist}.

In our solution, we adopt the AND-gate access policy {\bf AND}$_{m}^*$, which supports multiple values with wildcards *~\cite{zhang2014computationally}. An access policy {\it W} is a rule which returns either 0 or 1, given an attribute set {\it S}. That is, attribute set {\it S} satisfies policy {\it W} {\em if and only if} {\it W} evaluates to 1 on {\it S}. Note that the wildcard * in an AND-gate policy plays the role of a \say{don't care} value. Formally, given an attribute list {\it S} = [${\it S_1}$, ${\it S_2}$, ..., ${\it S_n}$] and an access policy {\it W} = [${{\it W_1}}$, ${\it W_2}$, ..., ${\it W_n}$] = $\bigwedge_{i\in I_W}$$W_i$, where $I_W$ is a subscript index set and $I_W$ = $\{{\it i |1\leq i \leq n, W_i \neq *} \}$, and we say that {\it S} $\models$ {\it W} if $S_i$ = $W_i$ or $W_i$ = * for all $1\leq i \leq n$ and {\it S} $\nvDash$ {\it W} otherwise~\cite{zhang2014computationally}. 

Here we present a concrete example. Suppose an data owner in the Computer and Information Sciences Department at the University of Delaware specifies access policy {\it W} to be [UD, PhD Student, Gender*] for accessing encrypted research meeting notes, and we have student Alice's attribute list $\it S_{Alice}$ = [UD, PhD Student, Female], and student Bob's attribute list $\it S_{Bob}$ = [UD, Master Student, Male]. As a result, Alice can access the corresponding encrypted research meeting notes, while Bob cannot because he is an Master student. Notice that the Gender* attribute indicates that either gender satisfies the access policy. 



\section{Architecture of Smart Contract based Multi Authority Access Control}
\label{sec:arch}

\begin{figure}[h]
\includegraphics[width=0.458\textwidth]{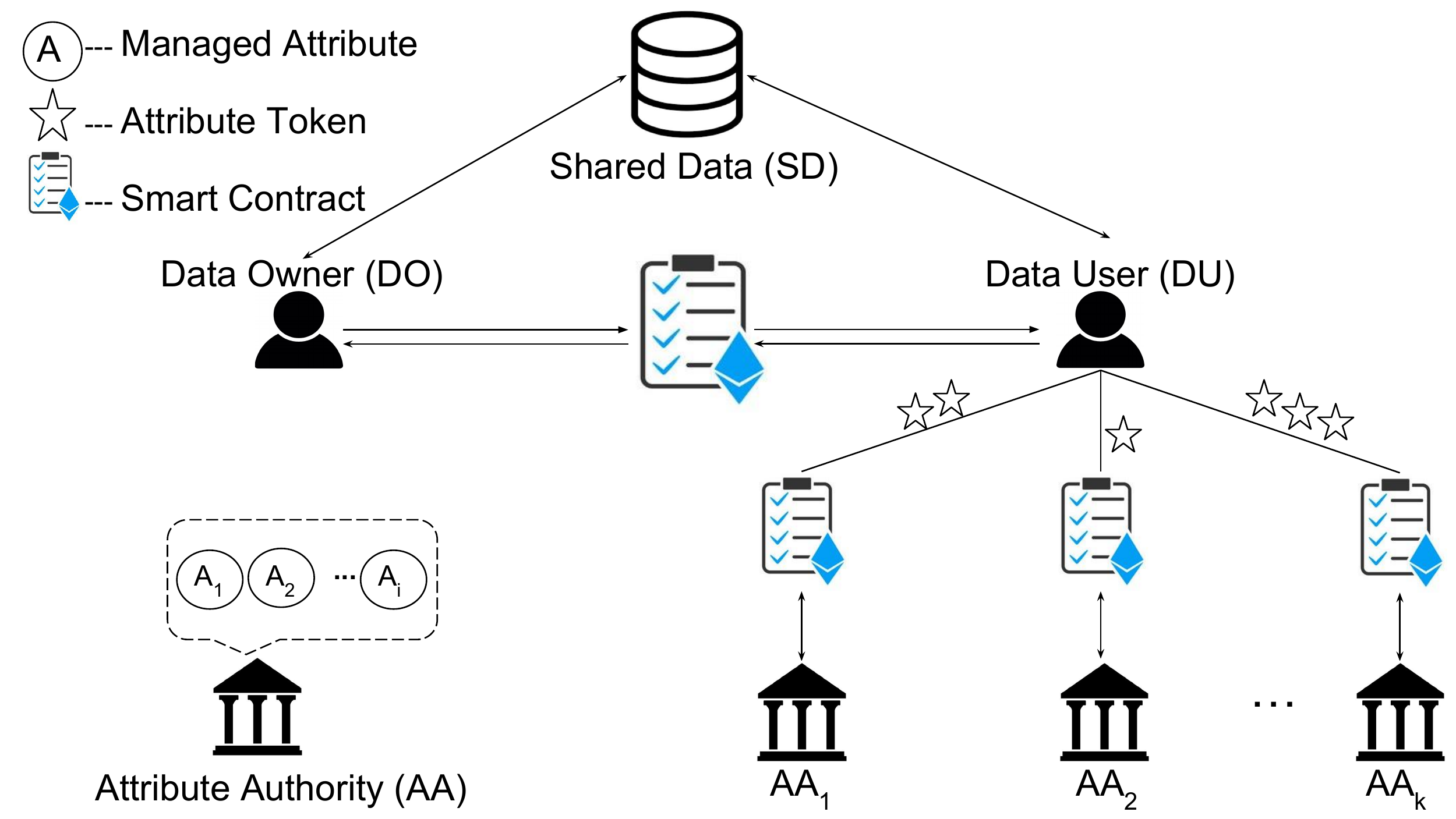}
\centering
\caption{Proposed system scheme.}
\label{fig:4}
\end{figure}

In this section, we describe the architecture of the proposed  multi-authority attribute-based access control scheme based on smart contracts.  By referring to Fig. \ref{fig:4}, we first enumerate the following entities that take part (either actively or passively) in the architecture.

\begin{figure*}[bt]
\centering
\includegraphics[width=0.88\textwidth]{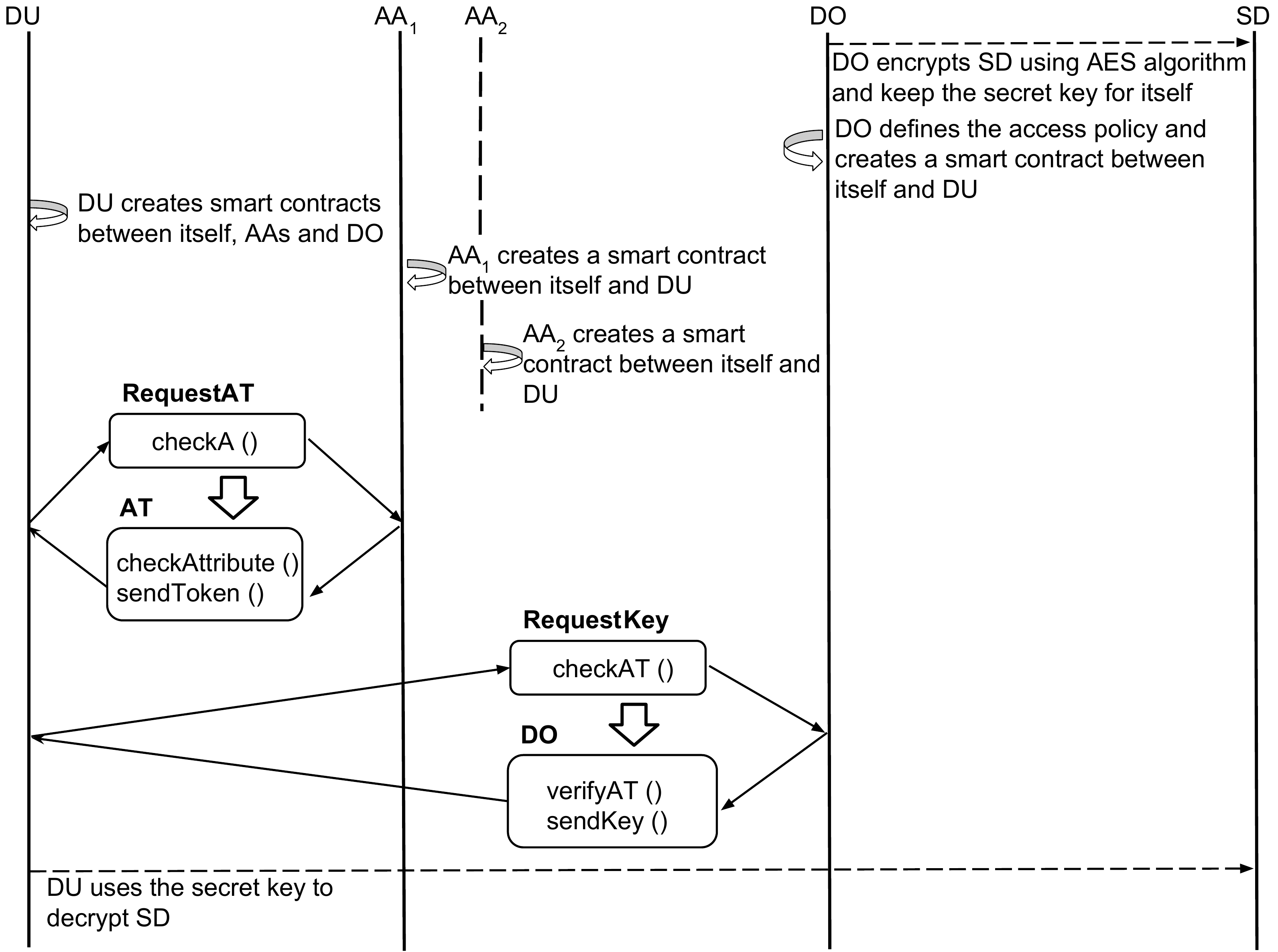}
\centering
\caption{Proposed workflow scenario.}
\label{fig:workflow}
\end{figure*}

\begin{itemize}
\item Data Owner (DO): A DO is an entity (e.g., person, organization, or process) who owns the data to be shared. A DO actively specifies access policies for the data it shares.

\item Data User (DU): A DU is an entity who wants to access data shared by DOs. A DU actively seeks access authorizations from DOs.

\item Shared Data (SD): An SD is a piece of data owned by a DO, and can be accessed passively by authorized DUs.

\item Attribute Token (AT): An AT is a credential representing an attribute that a DU possesses.  

\item Attribute Authority (AA): An AA is a pre-verified and authorized node in Ethereum who issues ATs to qualified DUs who possess the corresponding attributes.
\end{itemize}

In the proposed architecture,  after a DU has been validated for possessing a particular set of attributes by an AA, the AA will then issue the corresponding set of ATs to the DU. This validation process is carried out in the context of smart contracts. Consequently, satisfaction of the access policy associated with an SD is now represented by the collection of the corresponding ATs. 
Once a DU meets the access control policy imposed by the DO of the SD, the smart contract between the DU and the DO is executed for the DU to receive the AES secret key (which is encrypted with DU's public key) from the DO to access the SD.

Fig. \ref{fig:workflow} depicts a sample workflow scenario, where a DU, multiple AAs, and a DO communicate via smart contracts, as defined below, to perform multi-authority attribute-based access control. 

\begin{enumerate}
\item
AAs, DO, and DU register their respective EOA accounts with Ethereum, so that they may participate in Ethereum blockchain network. (Not shown in Fig. \ref{fig:workflow}.)

\item
Using a standard symmetric encryption algorithm, such as AES-256, the DO generates a secret key to encrypt the SD, and uploads the encrypted data to a shared database. In addition, the DO defines the access policy for the SD and creates smart contract {\bf DO} to be executed between itself and the DU.

\item A DU creates smart contract {\bf RequestAT} with each AA, which contains the function {\tt checkA} that requests validation of its attributes, and returns the corresponding ATs upon successful validation. The DU also creates smart contract {\bf RequestKey} with the DO, which contains function {\tt checkAT} that the DU holds enough ATs to send the AES secret key.

\item
AA$_1$ creates a smart contract {\bf AT} between itself and the DU, which contains function {\tt checkAttribute} for validating the DU's attributes, and the function {\tt sendToken} for granting ATs.
Similarly, AA$_2$ creates another smart contract {\bf AT} between itself and the DU. 

\item
Now, the DU wanting to access the SD executes its smart contract {\bf RequestAT} to request validation of its attributes by invoking function {\tt checkA}. 
In turns, {\tt checkA} triggers the execution of the smart contract {\bf AT} by invoking the {\tt checkAttribute} function of {\bf AT}. Upon successful validation, the {\tt sendToken} function is invoked to return the associated ATs to be saved in the ``balance'' of DU's EOA account.



\item After accumulating enough ``balance,'' the DU will executes  smart contract {\bf RequestKey} to invoke its function {\tt checkAT} for validating its ATs so as to obtain the AES secret key (encrypted with DU's public key) by triggering smart contract {\bf DO} and invoking its {\tt sendKey} function.





\item At the end, the DU decrypts the AES secret key (which was encrypted with its public key) with his private key, and uses the AES secret key to access the SD. 
\end{enumerate}




The smart contract created by DU between itself and AA: \\

\begin{mdframed}
{\texttt 
{contract {\textbf{RequestAT} \{ } {\\
\hspace*{0.5cm} function checkA(addressOfAA)\{ \\
\hspace*{1cm} {\textbf{AT} my\textunderscore at = {\textbf{AT}(addressOfAA);} \\ }
\hspace*{1cm} if (my\textunderscore at.checkAttribute() == true)\\
\hspace*{1.5cm}  return my\textunderscore at.sendToken(); \\ 
\hspace*{1cm}  return FAILURE; \\ 
\hspace*{0.5cm}  \} \\
 \hspace*{0.001cm} \}
  }
  }
} \end{mdframed} 
The smart contract created by DU between itself and DO: \\
\begin{mdframed}
{\texttt 
{contract {\textbf{RequestKey} \{ } {\\
 \hspace*{0.5cm} function checkAT(addressOfDO)\{ \\ {
 \hspace*{1cm}    {\textbf{DO} my\textunderscore ap = {\textbf{DO}(addressOfDO);} \\
  \hspace*{1cm}  if (my\textunderscore ap.verifyAT() == true) \\ 
  \hspace*{1.5cm} return my\textunderscore ap.sendKey(); \\
  \hspace*{1cm}  return FAILURE; \\ 
\hspace*{0.5cm}  \} \\
\hspace*{0.001cm}  \}
  }
}
}
}
}\end{mdframed}

The smart contract created by AA between itself and DU: \\

\begin{mdframed}
{\texttt 
{contract {\textbf{AT} is ERC20Interface, Owned\{ } {\\
 \hspace*{0.5cm} string public symbol; \\
 \hspace*{0.5cm} string public name; \\
 \hspace*{0.5cm} uint8 public decimals; \\ 
 \hspace*{0.5cm} mapping(address => uint) balances; \\ 
 \hspace*{0.5cm} mapping(uint256 => Data) CheckAttribute; \\ \\
 \hspace*{0.5cm} event Sendtoken(address from, address to,\\ \hspace*{0.5cm} uint tokens); \\ \\ 
\hspace*{0.5cm} struct Data\{ \\
\hspace*{1cm} uint256 AttributeID; \\
\hspace*{1cm} string attribute; \\
\hspace*{1cm} string approve; \\
     \hspace*{0.5cm}  \} \\ \\
\hspace*{0.5cm} function AttributeToken() public\{ \\
\hspace*{1cm} symbol = "UD"; \\
\hspace*{1cm} name = "UD Token"; \\
\hspace*{1cm} decimals = 0; \\
\hspace*{1cm} totalSupply = 100; \\
\hspace*{1cm} balances [addressOfAA] = totalSupply; \\
     \hspace*{0.5cm}  \} \\ 
 \hspace*{0.5cm} function checkAttribute(uint256 \\ \hspace*{0.5cm} AttributeID, string attribute, string \\ \hspace*{0.5cm} approve) public returns (bool  success)\{ \\
       \hspace*{1cm}  CheckAttribute[AttributeID] = \\ \hspace*{1cm} Data(AttributeID, attribute, approve); \\
       \hspace*{1cm} return true; \\
     \hspace*{0.5cm}  \} \\ \\
 \hspace*{0.5cm}     function sendToken(address to, uint tokens) \hspace*{0.5cm} public returns (bool  success)\{ \\
      \hspace*{1cm}   require(!frozenAccount[to]); \\
\hspace*{1cm}  emit Sendtoken(msg.sender, to, tokens); \\
\hspace*{1cm} return true; \\
     \hspace*{0.5cm}  \} \\
 \hspace*{0.01cm}   \} 
 }
}
}
\end{mdframed}





The smart contracts created by DO between itself and DU: \\
\begin{mdframed}
{\texttt 
{contract {\textbf{DO} \{ } {\\
\hspace*{0.5cm}  event Sendkey(address from, address to, \\\hspace*{0.5cm} bytes encryptedKey); \\ 
\hspace*{0.5cm}  event VerifyAT(address to, uint tokens,\\\hspace*{0.5cm} bytes approve); \\ \\
\hspace*{0.5cm} struct AESData\{ \\
\hspace*{1cm} bytes encryptedKey; \\
     \hspace*{0.5cm}  \} \\ \\
      \hspace*{0.5cm} function sendKey(address to, bytes \\ \hspace*{0.5cm} encryptedKey) public returns (bool \\\hspace*{0.5cm} success)\{ \\
       \hspace*{1cm}  emit Sendkey (msg.sender, to, \\\hspace*{1cm} encryptedKey); \\
       \hspace*{1cm} return true; \\
     \hspace*{0.5cm}  \} \\ \\
    \hspace*{0.5cm}  function verifyAT(address from, address \\ \hspace*{0.5cm} to, uint tokens, bytes approve) \\ \hspace*{0.5cm} public returns (bool success)\{ \\
    \hspace*{1cm} allowed [to][from] = tokens; \\
    \hspace*{1cm} require(balanceOf(to) >= \\ \hspace*{1cm} balanceOf(from)); \\
\hspace*{1cm}  emit VerifyAT(to, tokens, approve); \\
\hspace*{1cm} return true; \\
 \hspace*{0.5cm}  \} \\ 
 \hspace*{0.01cm}   \} 
  }
}
} 
\end{mdframed} 

Smart contracts provide the benefits of authentication, authorization, and audit as follows. First, when DO or DU executes a transaction, authentication is guaranteed since only the legitimate DO and DU accounts are able to initiate transactions. Second, the authorization is achieved by adopting ATs granted from multiple AAs to the DU if the DU has valid attributes, and later DO could check the ATs provided by the DU to determine if the DU satisfies the access policy or not. 
Third, by adopting the inherent design of blockchain, every transaction's record has been stored permanently and it's easy for people to audit the transaction information in future. In the end, each DU and DO has its own public and private keys, any attacker with no prior knowledge of the user's private key cannot decrypt the data send between DO and DU.

Overall, these smart contracts deploy the proposed scenarios as we described before. We implement the proof of concept prototype to illustrate the feasibility of our designed smart contracts and evaluate the experimental result in next section.

\section{Performance Evaluation and Security Analysis}
\label{sec:evaluation}

In this section, we evaluate the creation cost of smart contracts and the execution cost of their respective functions that facilitate multi-authority attribute-based access control via experiments on the Rinkeby Ethereum Testnet.
 
\subsection{Experiment Setting}

The smart contracts presented in the last section were programmed in Solidity~\cite{solidity} by using the official on-line IDE Remix.   In the month of October 2018, 1 ether $\approx$ 205 US\$. In our experiments, 1 gas = 1e9 ether, which is the minimum transaction cost. The lower the gas value, the longer time a transaction will be validated by miners, and vice versa.

\subsection{Results}

The costs of smart contract creations and their respective function executions are presented in Table \ref{tab:table1}. As we can observe from the table, the  one-time costs of smart contract creation for RequestAT, ResquestKey, and AT are 0.232, 0.208 and 0.359 US\$, respectively, based on the current price of ether. In comparison, the cost of function executions is relatively low.  Note that the execution costs of these functions vary based on the different input lengths such as the attribute information and other related data. 

\begin{table}[h]
\begin{tabular}{ 
|p{2.4cm}|p{1.2cm}|p{1.9cm}|p{1.2cm}|  }
 \hline
 \multicolumn{4}{|c|}{Experiment Results} \\
 \hline
 Contract/Function & Gas Used &Cost(ether)& USD (\$)\\
 \hline \hline
 RequestAT (C)   &1132452 & 0.001132452&  0.232\\ \hline
  RequestKey (C) &  1022644  & 0.001022644& 0.208\\ \hline
 AT (C) &  1752958  & 0.001752958   &0.359\\ \hline
 checkAttribute (F) &  394286  & 0.000394286   &0.080\\ \hline
 sendToken (F) &123680 & 0.000123680&  0.025\\ \hline
 verifyAT (F) & 356070  & 0.000356070   &0.073 \\ \hline
 sendKey (F) & 224520  & 0.000224520&0.046\\ \hline
 
\end{tabular}
\caption{\label{tab:table1}Cost of smart contracts and functions.}
\end{table}
\subsection{Security Analysis}
In our proposed scheme, we integrate the Ethereum smart contract, multi-authority attribute-based access control policy, and AES mechanism to provide a robust and privacy-preserving system. First, data owners fully control their personal data and there is no third-party authority to collect the information from the data owner. They have the right to distribute AES secret key to qualified data user. Second, in our solution, we utilize the multiple attribute authorities to avoid possible collusion and single point of failure. Since each attribute authority only responsible for certain number of attribute tokens, the attribute tokens reliability and availability is highly secured. Last, we use the Ethereum's blockchain platform to perform communications among different participants, which saves all the data records and transaction information. If later people like to trace back certain information, it will be easy to provide the data provenance records due to the character of blockchain technology.

\section{Related Work}

Several efforts have been done to provide the access control mechanism with the help of blockchain-based technologies. Maesa et al.~\cite{maesa2017blockchain} proposes a solution to publish  policies specifying the rights to access resources, and demonstrated a potential implementation based on deploying the eXtensible Access Control Markup Languag (XACML) \cite{xacml} on a Bitcoin blockchain network. Cruz et al.~\cite{cruz2018rbac} describes a role-based access control scheme using smart contracts, which is composed of two parts, the smart contracts to represent the trust and endorsement relationship and the challenge-response protocol to verify user identity. AI-Bassam et al.~\cite{al2017scpki} proposes the SCPKI (A Smart Contract-based PKI and Identity System), which is an PKI system based on a decentralized design using the web-of-trust model and a smart contract. The web-of-trust model enables an entity or authority in the system can verify attributes of another entity's identity (such as company name or domain name), as an alternative solution to the centralized authority identity verification model. Uchibeke et al.~\cite{uchibeke2018blockchain} proposed blockchain-based access control ecosystem which gives asset owners the sovereign right to manage access control for data sets and protect the data integrity. They use the Hyperledger composer tool to implement the smart contracts and other functions deployed on the blockchain network. 
Westerkamp et al.~\cite{westerkamp2018blockchaintoken} presents a blockchain-based supply chain traceability system using smart contracts. In their mechanism, each manufacturer defines the composition of products in the form of recipes, and each ingredient of the recipe is a non-fungible token which corresponds to a physical good. Overall, this system preserves the traceability of different product transformations.

In our work, we propose the multi-authority attribute-based access control scheme by using smart contract. Additionally, by utilizing the ERC-20 token standards and other functionalities provided by the solidity language, we map each unique attribute to attribute token to represent the requirement of access policy and other event-driven function to achieve the desired goal. 
\label{sec:related work}

\section{Conclusion}
\label{conclusion}

Our increasingly decentralized society motivates the need of multiple authorities to take part in attribute-based access control.
In this paper, we propose a multi-authority attribute-based access control mechanism. The interactions among data users, data owners, and multiple authorities are programmed into Ethereum smart contracts. We describe the architecture and operational workflow of multi-authority attribute-based access control. As a proof of concept, the scheme is programmed in Solidity and tested on the Rinkeby Ethereum Testnet. 





%

\bibliographystyle{ACM-Reference-Format}
\bibliography{sigproc}


\bibliographystyle{ACM-Reference-Format}
\bibliography{sigproc.bib} 

\end{document}